\documentclass[floatfix,twocolumn,aps,prd,superscriptaddress,longbibliography]{revtex4-1}

\usepackage{graphicx}% Include figure files
\usepackage{dcolumn}% Align table columns on decimal point
\usepackage{bm}% bold math
\usepackage{textcomp}
\usepackage[intlimits]{amsmath}
\usepackage{esint}
\usepackage{braket}
\usepackage{color}
\usepackage{soul} %for \hl and \st

\soulregister\ref{7}
\soulregister\eqref{7}
\soulregister\cite{7}
\soulregister\onlinecite{7}

\begin{document}
	
% TITLE
\title{Electrically pumped semiconductor laser with low spatial coherence and directional emission}

% AUTHOR & AFFILATION
\author{Kyungduk Kim}
\affiliation{Department of Applied Physics, Yale University, New Haven, Connecticut 06520, USA}%
\author{Stefan Bittner}
\affiliation{Department of Applied Physics, Yale University, New Haven, Connecticut 06520, USA}%
\author{Yongquan Zeng}
\affiliation{Center for OptoElectronics and Biophotonics, School of Electrical and Electronic Engineering and the Photonics Institute, Nanyang Technological University, 639798 Singapore}%
\author{Seng Fatt Liew}
\affiliation{Department of Applied Physics, Yale University, New Haven, Connecticut 06520, USA}%
\author{Qijie Wang}
\affiliation{Center for OptoElectronics and Biophotonics, School of Electrical and Electronic Engineering and the Photonics Institute, Nanyang Technological University, 639798 Singapore}%
\author{Hui Cao}
\email{hui.cao@yale.edu}
\affiliation{Department of Applied Physics, Yale University, New Haven, Connecticut 06520, USA}%

% ABSTRACT
\begin{abstract}
We design and fabricate an on-chip laser source that produces a directional beam with low spatial coherence. The lasing modes are based on the axial orbit in a stable cavity and have good directionality. To reduce the spatial coherence of emission, the number of transverse lasing modes is maximized by fine-tuning the cavity geometry. Decoherence is reached in a few nanoseconds. Such rapid decoherence will facilitate applications in ultrafast speckle-free full-field imaging.
\end{abstract}

\maketitle % End of title

%  MAIN TEXT

The high spatial coherence of conventional lasers can introduce coherent artifacts due to uncontrolled diffraction, reflection and optical aberration. A common example is the speckle formed by the interference of coherent waves with random phase differences~\cite{goodman2007speckle, dainty2013laser}. Speckle noise is detrimental to full-field imaging applications such as displays~\cite{chellappan2010laser}, microscopy, optical coherence tomography, and holography~\cite{bianco2018strategies}. It also poses as a problem for laser-based applications like material processing, photolithography~\cite{noordman2009speckle}, and optical trapping of particles~\cite{neuman2004optical}.
    
Various approaches to mitigate speckle artifacts have been developed. A traditional method is to average over many independent speckle patterns generated by a moving diffuser~\cite{lowenthal1971speckle,kubota2010very}, colloidal solution~\cite{redding2013low}, or fast scanning micromirrors~\cite{akram2010laser}. However, the generation of a series of uncorrelated speckle patterns is time-consuming and limited by the mechanical speed. A more efficient approach is to design a multimode laser that generates spatially incoherent emission, thus directly suppressing speckle formation~\cite{cao2019complex}. Low spatial coherence necessitates lasing in numerous distinct spatial modes with independent oscillation phases. For example, a degenerate cavity~\cite{nixon2013efficient, chriki2015manipulating, knitter2016coherence} allows a large number of transverse modes to lase, but the setup is bulky and hard to align. Complex lasers with compact size such as random lasers~\cite{gouedard1993generation, redding2011spatial, redding2012speckle, hokr2016narrow} have low spatial coherence and high photon degeneracy, but are mostly on optically pumped. For speckle-free imaging applications, wave-chaotic semiconductor microlasers~\cite{redding2015low} have the advantages of electrical pumping and high internal quantum efficiency. However, disordered or wave-chaotic cavity lasers typically have no preferential emission direction, and the poor collection efficiency greatly reduces their external quantum efficiency. Our goal is creating an electrically pumped multimode semiconductor microlaser without disordered or wave-chaotic cavity to combine low spatial coherence and directional emission.
    
Moreover, the speed of speckle suppression is crucial for imaging applications. For instance, time-resolved optical imaging to observe fast dynamics requires speckle-free image acquisition with a short integration time, so the oscillation phases of different spatial lasing modes must completely decorrelate during the integration time to attain decoherence. The finite linewidth $\Delta\nu$ of individual lasing modes leads to their decoherence on a time scale of $1/\Delta\nu$. The frequency difference between different lasing modes can lead to even faster decoherence. For example, the emission from many random lasing modes with distinct frequencies exhibits low spatial coherence already within ten nanoseconds~\cite{mermillod2013time}. The decoherence time was measured for a solid-state degenerate cavity laser~\cite{chriki2018spatiotemporal}. The intensity contrast of laser speckle is reduced by the dephasing between different longitudinal mode groups in tens of nanoseconds, but complete decoherence requires a few microseconds due to the small frequency spacing between transverse modes. We aim to further shorten the decoherence time by utilizing the larger mode spacings in a semiconductor microlaser. 

    % FIGURE 1
	\begin{figure}[t]
		\centering
		\includegraphics[width=\linewidth]{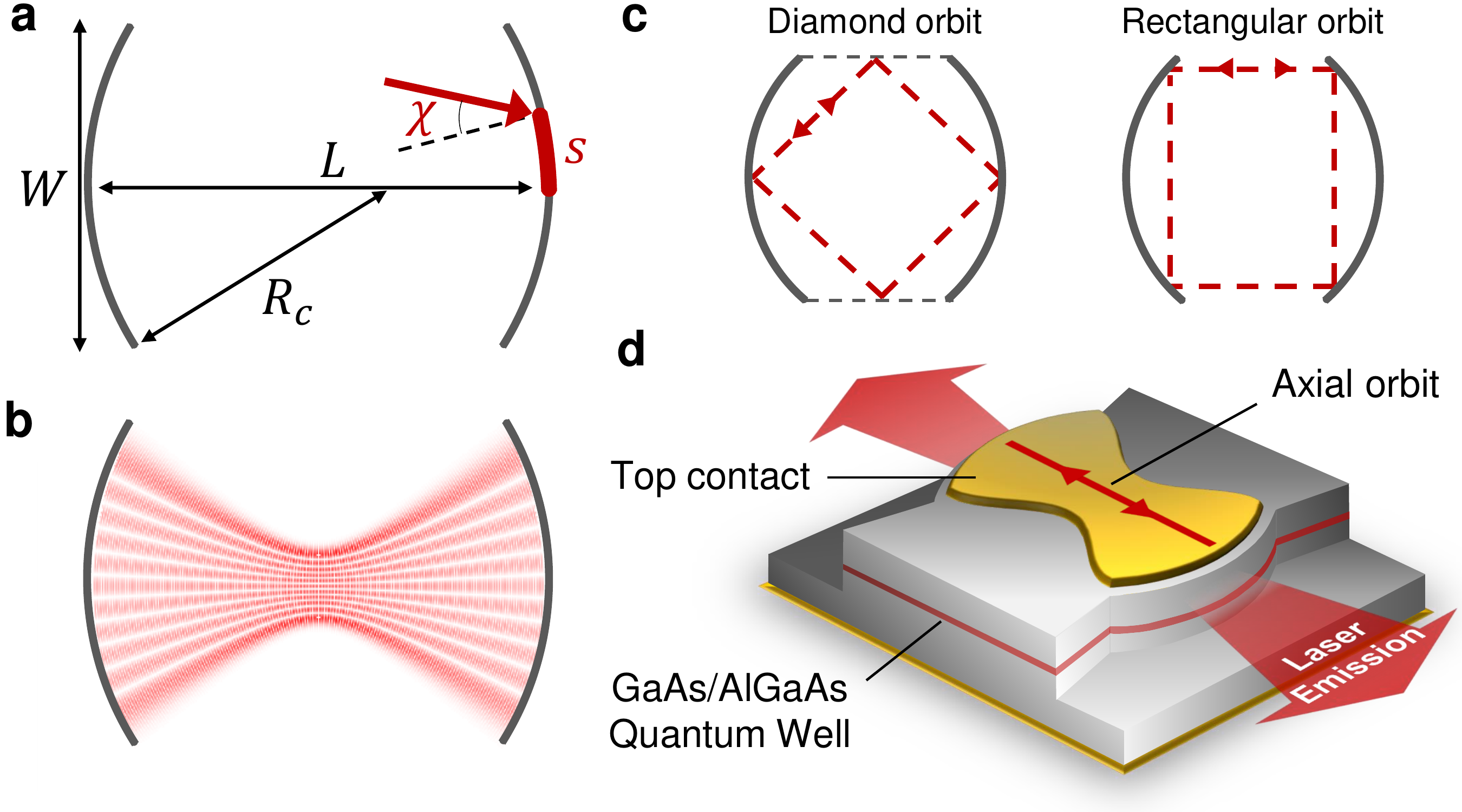}
		\caption{\textbf{Schematic of on-chip stable cavity laser.}
			\textbf{a}, 2D symmetric stable cavity defined by two concave circular mirrors with radius of curvature $R_c$, distance $L$ and cavity width $W$. Rays impinging on the cavity boundary are described by coordinates $(s, \chi)$, where $s$ is the coordinate along the curved boundary, and $\chi$ is the angle of incidence with respect to the surface normal.
			\textbf{b}, The spatial intensity profile of a high-order transverse mode in a stable cavity.
			\textbf{c}, Non-axial orbits that lead to non-directional emission.
			\textbf{d}, Three-dimensional sketch of the on-chip stable cavity with directional emission. The lasing modes are based on the axial orbit and thus have directional emission. The shape of the top metal contact matches the spatial profile of high-order transverse modes to ensure their spatial overlap with gain.}
		\label{figure1}
	\end{figure}

% FIGURE 1 - MAIN TEXT
	
Here we design an electrically-pumped chip-scale semiconductor laser with spatially incoherent and directional emission. The emission from conventional broad-area semiconductor lasers with flat end mirrors exhibits a good directionality. However, lasing occurs only in a few transverse modes since the high-order transverse modes have large divergence angles and hence experience severe losses \cite{Lang1991,Hartmann2017}. To lower the spatial coherence, we need to increase the number of transverse lasing modes. With curved end mirrors, the losses of high-order transverse modes can be reduced. We consider two-dimensional (2D) symmetric cavities with two circular concave mirrors with radius of curvature $R_c$ as shown in Fig.~\ref{figure1}a. The mirrors are separated by the cavity length $L$. The geometry of the cavity is determined by the parameter $g=1-L/R_c$, which is known as cavity stability parameter. If $R_c$ is larger than $L/2$, or $g$ is within the range $(-1, 1)$, the cavity is called stable in the sense that rays starting near the axial orbit stay close to it and will remain inside the cavity~\cite{siegman1986lasers}. In the paraxial limit, the resonances in the stable cavity are described by Hermite-Gaussian modes, which have different transverse profiles depending on the transverse mode number $m$. Figure~\ref{figure1}b shows the spatial intensity profile of a high-order transverse modes with $m=10$.

Reducing the speckle contrast to below the level of human perception $\simeq 0.03$ requires 1000 transverse modes to lase simultaneously and independently~\cite{roelandt2014human, geri2012perceptual}. Previous designs of stable cavity semiconductor lasers with curved facets~\cite{biellak1995lateral, fukushima2002ring, fukushima2012lowest} exhibited less than $10$ transverse lasing modes. The challenge is to increase the number of transverse lasing modes by two orders of magnitude. To accommodate higher order transverse modes, we increase the cavity width $W$. However, modes based on non-axial orbits like those in Fig.~\ref{figure1}c can appear in wide cavities, yielding non-directional emission. To ensure directional emission, all lasing modes must be based on the axial orbit. We eliminate the reflecting surfaces at the lateral sides, to suppress the non-axial modes based on the periodic orbits with bounces from the sidewalls, such as the diamond orbit~\cite{fukushima2002ring, fukushima2012lowest}. In addition,  we set $W = L / \sqrt{2}$ to avoid the rectangle orbits in the stable cavity. A schematic of our design is shown in Fig.~\ref{figure1}d. The top metal contact for current injection is shaped to match the profile of high-order transverse modes, to ensure their spatial overlap with the gain. 

% FIGURE 2 - MAIN TEXT
To maximize the number of high-$Q$ transverse modes, we optimize the cavity shape by fine tuning $R_c$ while keeping $L$ and $W$ fixed. We numerically calculate the passive cavity modes using the finite element method (COMSOL). A 2D cavity with $L=20~\mu$m is simulated. The refractive index of the cavity $n=3.37$ corresponds to the effective refractive index of the vertically guided mode in the GaAs wafer used in the experiment. Transverse-electric (TE) polarization (electric field parallel to the cavity plane) is considered since GaAs quantum wells have higher gain for this polarization and the lasing modes are TE polarized.
    
    % FIGURE 2
	\begin{figure}[t]
		\centering
		\includegraphics[width=\linewidth]{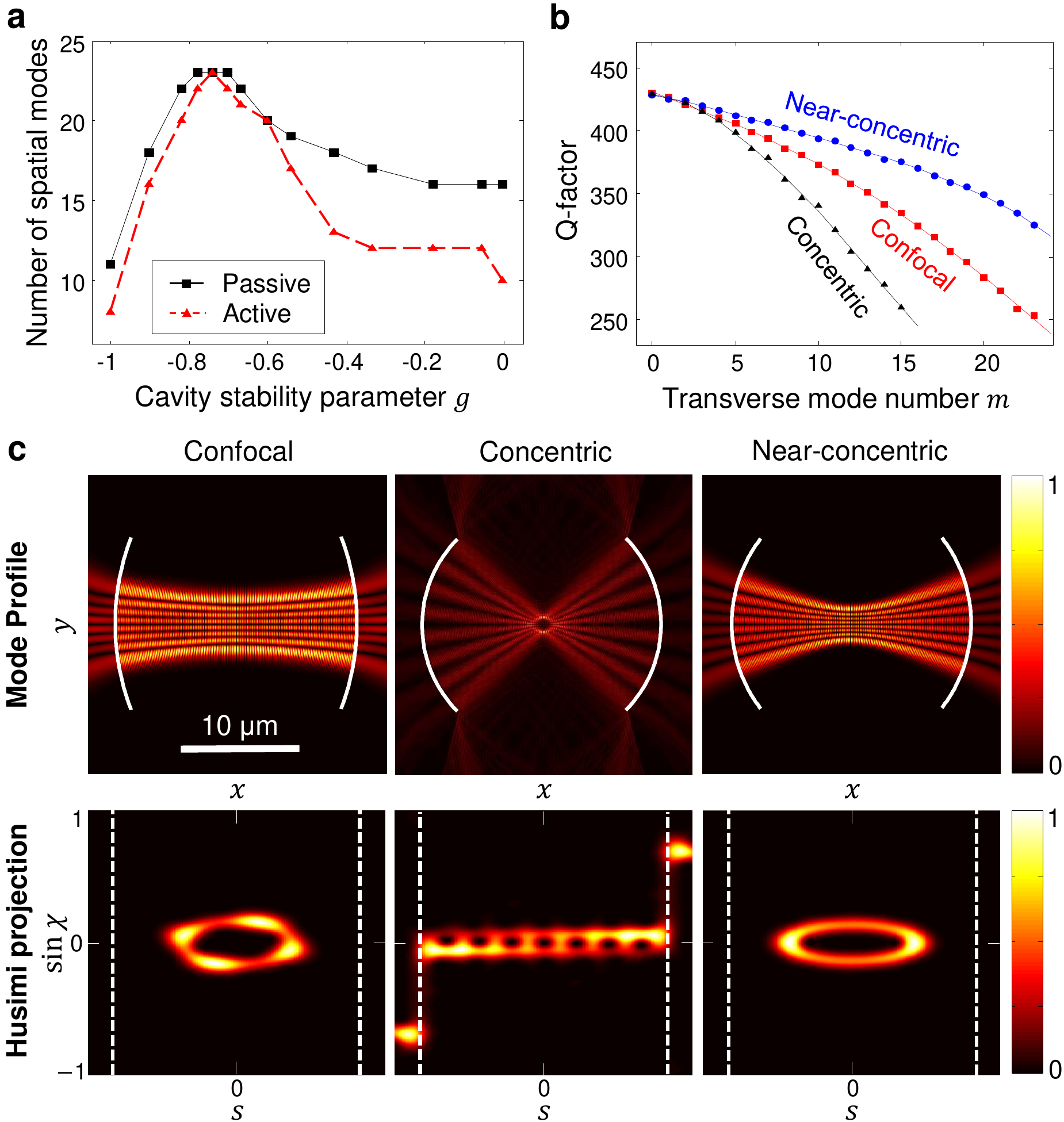}
		\caption{\textbf{Fine tuning of the cavity geometry to maximize the number of transverse modes.}
			\textbf{a}, The number of high-$Q$ passive modes (black squares) and the number of lasing modes (red triangles) that are based on axial orbit and exhibit distinct transverse profiles.
			\textbf{b}, Dependence of quality factor $Q$ on transverse mode number $m$ for the optimized near-concentric ($g=-0.74$), confocal ($g=0$), and concentric ($g=-1$) cavities.
			\textbf{c}, Calculated spatial distributions of field amplitude (top) and corresponding Husimi projections (bottom) for high-order transverse modes ($m=7$) in cavities of different $g$. The white solid lines in the top row represent the curved cavity facets, and the dashed lines in bottom row mark the endpoints of the facets.}
		\label{figure2}
	\end{figure}

Degenerate cavities with conventional mirrors can support transverse modes with nearly-degenerate $Q$-factors thanks to their self-imaging property~\cite{arnaud1969degenerate,gigan2005image}. As an example of a degenerate cavity we consider the confocal geometry ($g=0$). For the on-chip design with dielectric interfaces as mirrors, however, the $Q$-factor decreases significantly as the transverse mode number $m$ increases as shown in Fig.~\ref{figure2}b. Figure~\ref{figure2}c shows a typical mode laterally confined to the cavity axis, resulting in negligible diffraction loss. However, its Husimi projection~\cite{hentschel2003husimi}, which visualizes the angle of incidence of wave components on different parts of the cavity boundary, features high-intensity spots at nonzero incident angle $\chi$. As $m$ increases, wave components with increasingly higher incident angles appear. Thus high-order transverse modes in the confocal cavity experience higher loss since the reflectivity at a dielectric-air interface decreases with increasing $\chi$ for TE-polarized light, making the confocal cavity unsuitable for multimode lasing.
   				
To solve this problem we consider the concentric cavity ($g=-1$). Since the concentric mirrors are part of a circle, any ray passing through the cavity center hits the boundaries perpendicularly. Indeed the Husimi projection in Fig.~\ref{figure2}c is strongly localized at $\chi=0$ and thus the angle-dependent reflectance is an insignificant loss mechanism. However, as the mode profile exhibits a large divergence, light leaks out via diffraction from the endpoints of the facets. These losses are evident in the Husimi projection from the high-intensity spots just outside the cavity facet. Since the higher order transverse modes experience stronger diffraction loss, the $Q$-factor decreases even more quickly with $m$ than for the confocal case as seen in Fig.~\ref{figure2}b.

We gradually vary $g$ from -1 to 0 in search for the optimal geometry that supports the largest number of high-$Q$ transverse modes. Fig.~\ref{figure2}a shows the number of transverse modes, that are based on the axial orbit and have $Q$-factors exceeding $0.8$ times the maximal $Q$-factor, as a function of $g$. The optimal geometry $g=-0.74$ is near concentric. A slight deviation from the concentric shape makes the mode profiles laterally localized to the cavity axis (see Fig.~\ref{figure2}c). Moreover, the Husimi projection shows high-intensity spots centered at $\chi=0$, which indicates most wave components have almost normal incidence on the cavity facet. Therefore, the near-concentric geometry minimizes both losses from angle-dependent reflectance and diffraction, resulting in the slowest decrease of $Q$ with $m$. As the number of transverse modes scales linearly with the width $W$ of the cavity when keeping the ratio $W/L$ fixed, we can apply this optimization result to the larger cavities used in experiments (see Supplementary).
    
The above optimization is based on the passive cavity modes. Gain competition can limit the number of lasing modes additionally. In order to quantify the effect of gain competition, we calculate the number of lasing modes at steady state~\cite{ge2010steady,cerjan2016controlling} (see Materials and Methods). The red curve in Fig.~\ref{figure2}a represents the number of different transverse lasing modes at a pump level of two times above the lasing threshold. In the confocal cavity, the number of lasing modes based on axial orbit is notably smaller than the number of high-$Q$ passive modes due to the existence of non-axial modes with higher $Q$ that lase first and saturate the gain for the axial modes (see Supplementary). For the optimized near-concentric cavity, most of the passive transverse modes with high $Q$ can lase, indicating gain competition is insignificant.
    
	% FIGURE 3
	\begin{figure}[b]
		\centering
		\includegraphics[width=\linewidth]{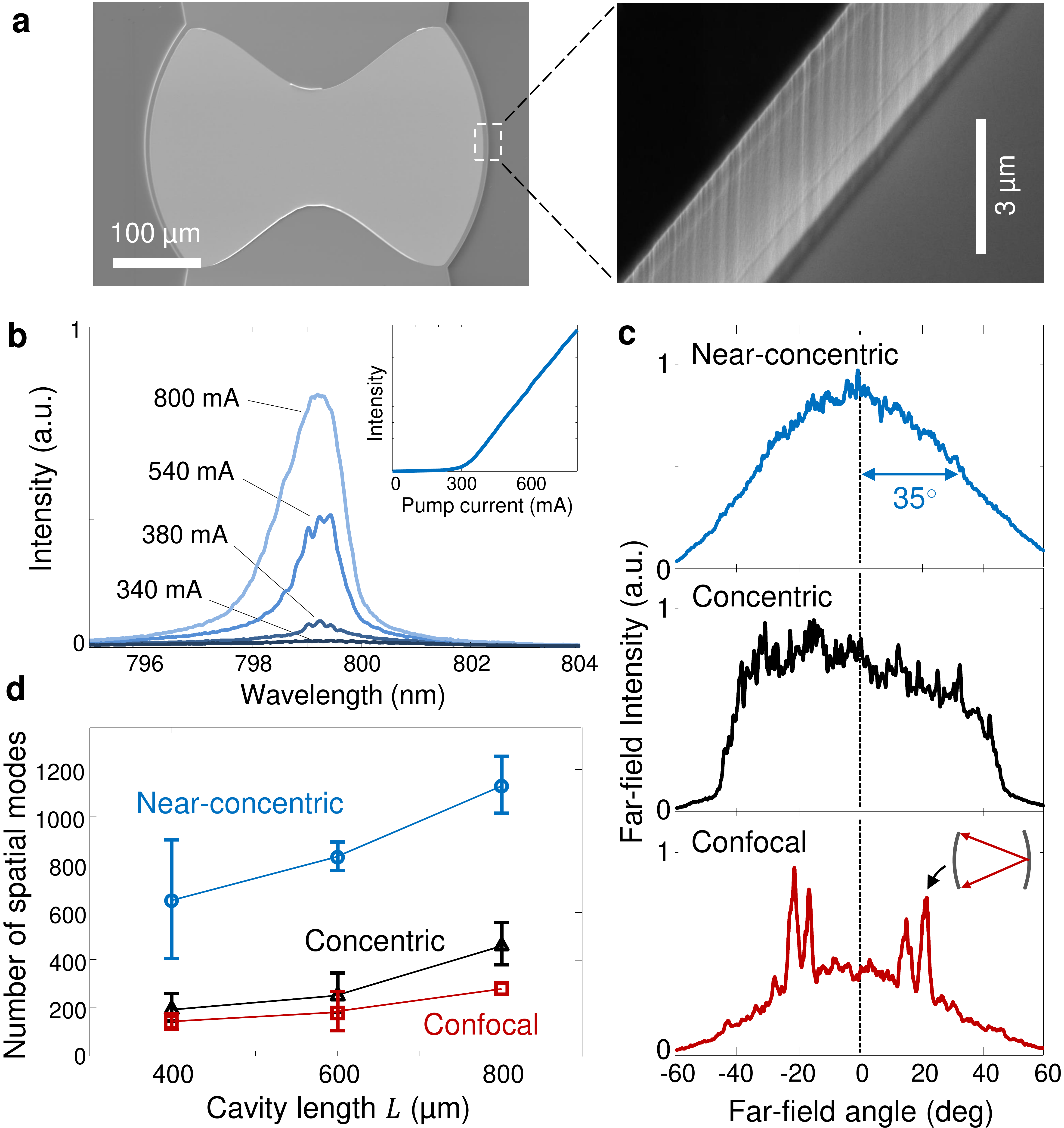}
		\caption{\textbf{Lasing characteristics of the on-chip stable cavities.}
		\textbf{a}, SEM images of a fabricated near-concentric ($g=-0.74$) cavity of length $L = 400$ $\mu$m. The etched facet is vertical and smooth.
		\textbf{b}, Emission spectra at different pump currents of a near-concentric cavity with $L=400$ $\mu$m. The L-I curve is plotted in the inset.
		\textbf{c}, Far-field intensity patterns of laser emission from three cavities with $g=-0.74, -1,$ and $0$. The V-shaped orbit, which contributes to the sharp peaks in the far-field pattern of the confocal cavity, is drawn in the inset.
		\textbf{d}, The number of transverse lasing modes in near-concentric ($g=-0.74$), concentric ($g=-1$), and confocal ($g=0$) cavities with different length $L$. The error bars indicate the variation between different cavities of the same $g$ and $L$.
		}
		\label{figure3}
	\end{figure}
	
	% FIGURE 3 - MAIN TEXT
The experimental results for the on-chip stable cavity lasers are presented in Fig.~\ref{figure3}. The cavities are fabricated with photolithography followed by reactive ion etching from a commercial GaAs/AlGaAs quantum well epiwafer. The scanning electron microscope (SEM) images in Fig.~\ref{figure3}a show that the etched facets, which serve as curved end mirrors, are smooth and vertical. The fabricated sample is mounted on a copper block and a tungsten needle is placed on the top gold contact for current injection. Lasing is observed at room temperature with electrical pumping for all the tested cavities with different sizes and shapes. To reduce heating, the current pulses are $2~\mu$s-long with $10$~Hz repetition rate. The emission is collected by an objective lens (NA = 0.4) and coupled into a spectrometer. Figure~\ref{figure3}b shows the emission spectra from an optimized near-concentric cavity ($g=-0.74$) at different pump currents. A typical spectrum consists of many closely-spaced narrow peaks, indicating simultaneous lasing of many modes. More lasing peaks appear at higher pump currents, and they merge to a smooth, broad spectrum. The L-I curve for a $L=400~\mu$m-long cavity (inset of Fig.~\ref{figure3}b) shows the lasing threshold is $360$~mA. The threshold current density is inversely proportional to the cavity length $L$ (not shown), as expected since the $Q$-factors increase linearly with $L$. There was no significant difference between the lasing thresholds for cavities with the same $L$ but different $g$.
	
To investigate the emission directionality, we measure the far-field emission patterns at a pump current two times above the lasing threshold. Figure~\ref{figure3}c shows the far-field patterns for three cavity shapes. For a near-concentric cavity ($g=-0.74$), a directional output beam with a divergence angle (half width at half maximum) of 35$^\circ$ is observed. The concentric cavity ($g=-1$) shows a flat-top far-field pattern with sharp edges. This pattern is attributed to the broad angular divergence of modes in the concentric cavity. In contrast, the far-field pattern of the confocal cavity features sharp peaks on top of a broad background. The sharp peaks originate from lasing modes based on a V-shaped, non-axial orbit (see inset and Supplementary).
	
\begin{figure}[t]
		\centering
		\includegraphics[width=\linewidth]{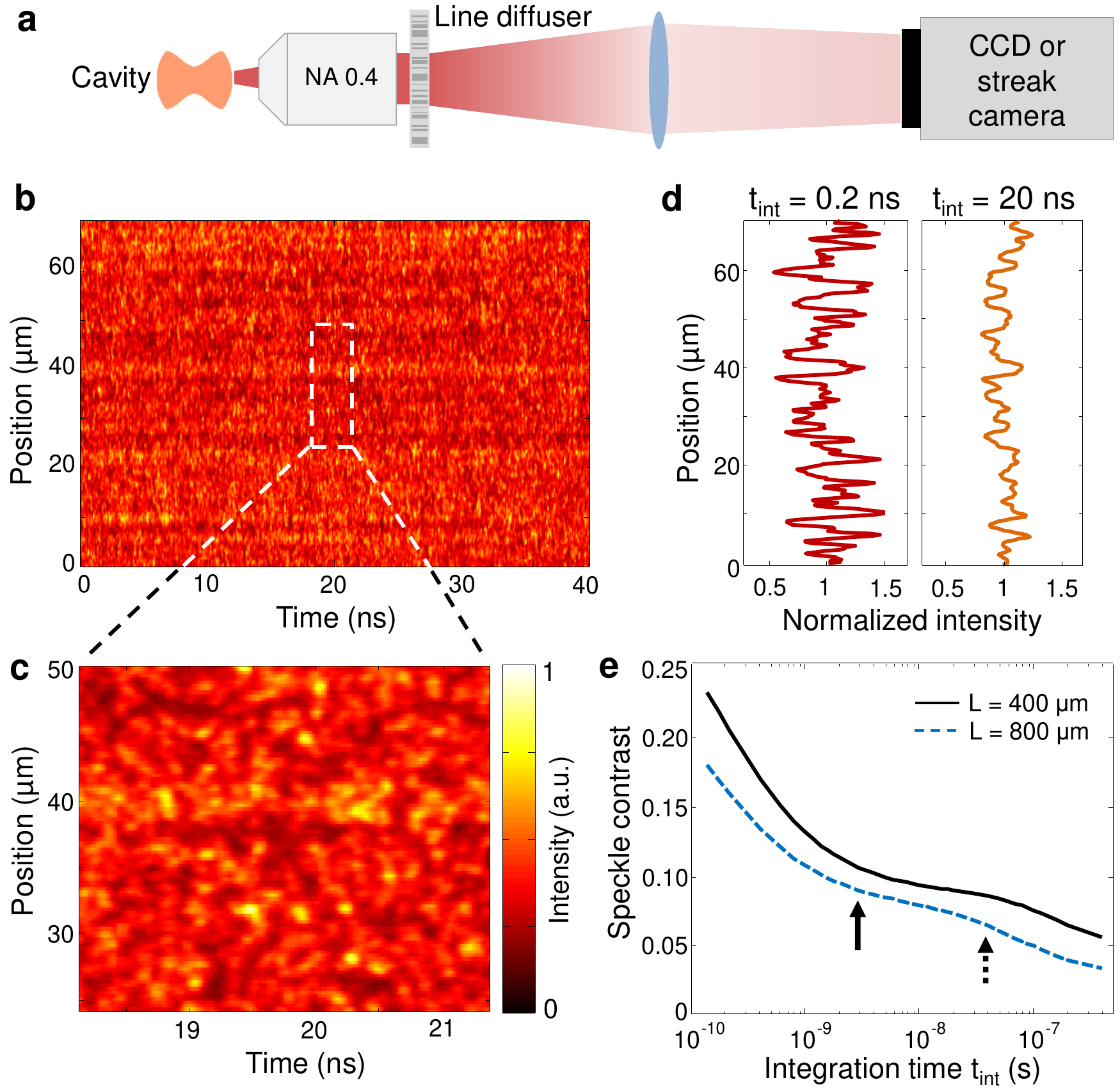}
		\caption{\textbf{Decoherence time for a near-concentric cavity laser.}
		    \textbf{a}, Schematic of the setup that measures speckle patterns with a CCD or streak camera.
			\textbf{b}, The spatio-temporal profile of the far-field speckle pattern from a laser with $g=-0.74$ and $L=400~\mu$m.
			\textbf{c}, Magnification of the speckle pattern revealing fast temporal evolution of the speckle grains. 			
			\textbf{d}, Time-integrated speckle patterns for integration times $0.2$~ns and $20$~ns, exhibiting an intensity contrast of $C = 0.21$ and $0.098$, respectively.
			\textbf{e}, Dependence of speckle contrast on the integration time $t_\mathrm{int}$, featuring two kinks at the integration times of a few nanoseconds (solid arrow) and several tens of nanoseconds (dotted arrow).}
		\label{figure4}
	\end{figure}

We characterize the spatial coherence of the laser emission from the cavities of different shapes. The emission is coherent in the direction normal to the wafer since the sample has only one index-guided mode in the vertical direction. To measure the coherence of emission in the horizontal direction (parallel to the wafer), we create speckle patterns with a line diffuser that scatters light only in the horizontal direction. A CCD camera records the far-field speckle intensity pattern generated by laser emission from a single $2~\mu$s-long pump pulse (see Fig.~\ref{figure4}a). In order to quantify the spatial coherence, we calculate the speckle contrast defined as $C=\sigma_I/\langle I \rangle$, where $\sigma_{I}$ and $\langle I \rangle$ are the standard deviation and mean of the speckle intensity, respectively. $M=1/C^2$ gives the effective number of distinct transverse lasing modes~\cite{goodman2007speckle}. Figure~\ref{figure3}d gives the values of $M$ for cavities with different $g$ and $L$, measured at two times of the lasing threshold. The number of transverse lasing modes is the largest for the near-concentric cavity ($g=-0.74$). With the ratio $W/L$ fixed, the number of transverse modes increases with $L$ since a wider cavity supports more transverse modes. For the $L=800~\mu$m near-concentric cavity ($g=-0.74$), about $1,000$ different transverse modes lase, and their combined emission reduces the speckle contrast to about $0.03$ which is the level of below human perception \cite{roelandt2014human,geri2012perceptual}.

	% FIGURE 4

% FIGURE 4
To examine the applicability of the optimized laser for ultrafast speckle-free imaging, we determine how fast decoherence of the emission occurs. We use a streak camera to measure the time-resolved speckle patterns with a temporal resolution of about $60$~ps in a setup sketched in Fig.~\ref{figure4}a. Figure~\ref{figure4}b shows the spatio-temporal evolution of the measured far-field speckle pattern of a near-concentric cavity laser ($g=-0.74$). The magnification in Fig.~\ref{figure4}c reveals rapid spatial and temporal variations of the intensity pattern. To quantify the coherence time of the emission, we calculate the contrast of speckle patterns as a function of the integration time. As shown in Fig.~\ref{figure4}d, for a short integration time of $t_\mathrm{int} = 0.2$~ns, the speckle has a notable contrast of $\sim 0.2$. As the integration time increases, the speckle contrast drops quickly. Figure~\ref{figure4}e summarizes the reduction of the speckle contrast for $t_\mathrm{int}$ from $100$~ps to $500$~ns. The $L=800~\mu$m-long cavity laser features lower speckle contrast than the $L=400~\mu$m-long cavity laser for all integration times. After a rapid drop, the contrast starts to saturate, exhibiting a kink at a few nanoseconds (indicated by the solid arrow). A second kink (indicated by a dotted arrow) follows at several tens of nanoseconds after which the speckle contrast further declines.
	
The time scale of the speckle contrast reduction is related to the frequency differences of lasing modes when their linewidths are smaller than their frequency spacings. When the integration time $t_\mathrm{int}$ is shorter than the inverse frequency spacing of two modes, their temporal beating results in a visible interference pattern that oscillates in time. For an integration time longer than their beating period, the time-varying interference pattern is averaged out, hence the intensity contrast of the speckle pattern created by these two modes is reduced. With increasing integration time, more and more lasing modes become incoherent, as their frequency spacings exceed $1/t_\mathrm{int}$, and  the speckle contrast continues dropping. Once $t_\mathrm{int}$ is long enough to average out the beating of even the closest pairs of lasing modes, the speckle contrast cannot reduce further. The average frequency spacing between adjacent modes is estimated as several hundred MHz in our cavities (See supplementary), whereas the typical linewidth of semiconductor lasers ($10$--$100$~MHz) is smaller than the frequency spacing. Thus the integration time needed for contrast reduction is determined by the mode spacing and estimated to be a few nanoseconds, which matches the experimental observations. The additional reduction of the speckle contrast at a few ten nanoseconds is attributed to a thermally-induced change of lasing modes~\cite{bittner2018suppressing} (See supplementary). When the lasing modes change, the output emission patterns change as well and their superposition further reduces the speckle contrast.
	
	% DISCUSSION
In summary, we demonstrate directional emission, low spatial coherence and ultrashort decoherence time in a compact electrically-pumped semiconductor laser. By optimizing the shape of an on-chip near-concentric cavity, we maximize the number of transverse lasing modes and thus greatly suppress speckle formation. Low speckle contrast is obtained even with an integration time of a few nanoseconds. Such short decoherence time enables ultrafast speckle-free full-field imaging. Finally, we compare this work to the previous demonstration of spatially-incoherent non-modal emission from a broad-area vertical-cavity surface-emitting laser \cite{peeters2005spatial}. By carefully adjusting the pump conditions, the cavity is constantly modified by thermal effects, which disrupts the formation of lasing modes, leading to spatially incoherent emission \cite{mandre2008evolution, craggs2009thermally}. Our approach does not rely on thermal effects, and the decoherence time is two orders of magnitude shorter. Furthermore, our method does not utilize any transient process, thus it is applicable to steady-state lasing. With better thermal management, our laser may operate under constant pumping, emitting a continuous wave of low spatial coherence. 
    	
\section*{Acknowledgments}
We thank N.~Davidson, R.~Chriki, and A.~D.~Stone for fruitful discussions. This work conducted at Yale University is supported by the Air Force Office of Scientific Research (AFOSR) under grant FA9550-16-1-0416, and by the Office of Naval Research (ONR) with MURI grant N00014-13-1-0649. For the work at Nanyang Technological University, funding support is acknowledged from the the Ministry of Education, Singapore grant (MOE2016-T2-1-128, MOE2016-T2-2-159) and National Research Foundation, Competitive Research Program (NRF-CRP18-2017-02).

\bigskip\bigskip\bigskip
%%%%% Supplementary %%%%%

\section*{SUPPLEMENTARY MATERIAL}

	% Materials and methods
	\subsection*{Materials and methods}
	
	% FIGURE 1
	\begin{figure}[b]
		\centering
		\includegraphics[width=\linewidth]{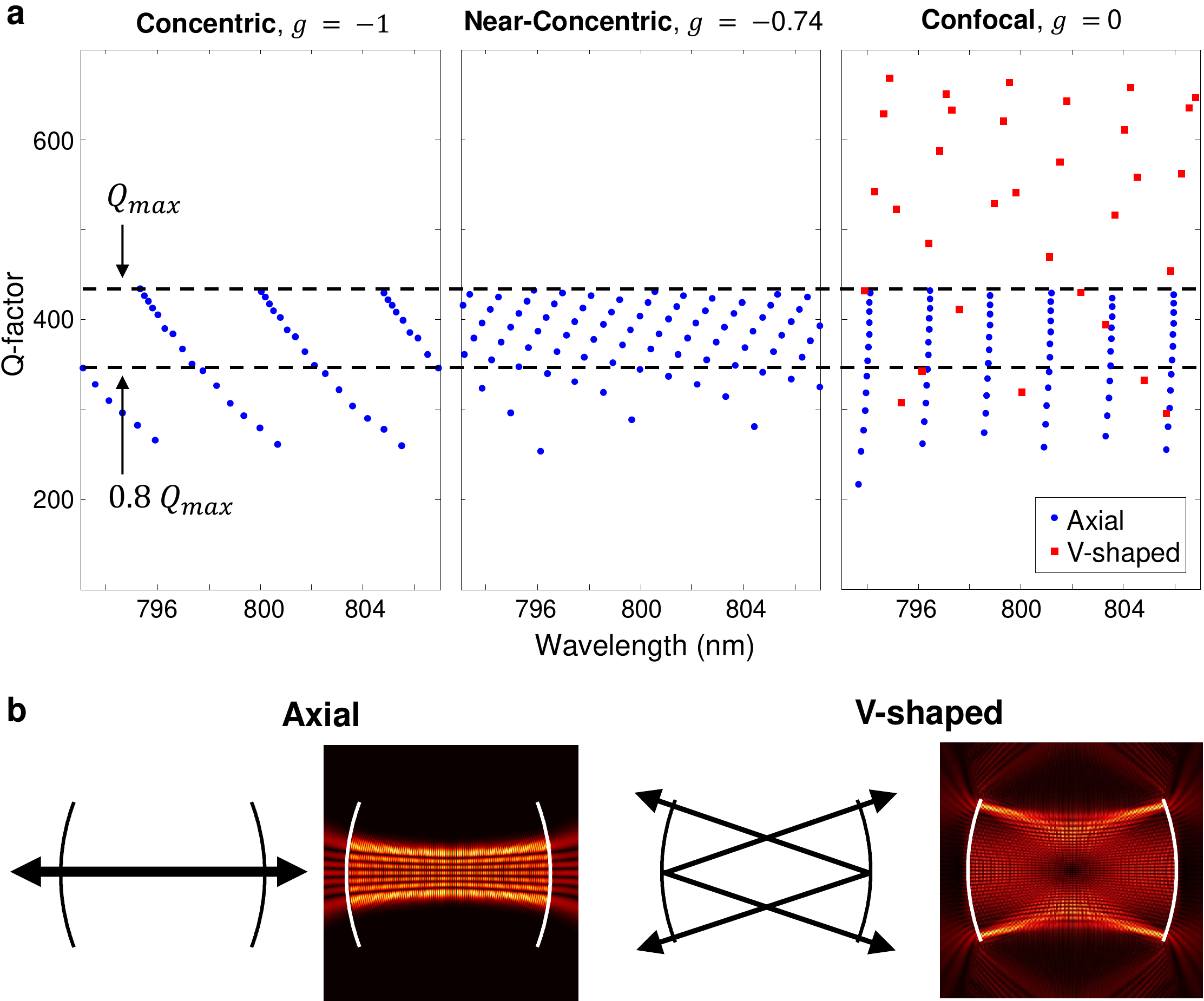}
		\caption{
			\textbf{High-$Q$ modes in passive cavities of different geometries.}
			\textbf{a}, Calculated quality factors and wavelengths of resonances in concentric ($g=-1$), near-concentric ($g=-0.74$), and confocal ($g=0$) cavities. The upper dashed lines indicate the maximum $Q$ factor $Q_{max}$. The lower dashed line is $0.8 \, Q_{max}$, the lower limit of high-$Q$ modes in our consideration. Blue dots are modes based on the axial orbit and red squares are modes on a non-axial orbit. 
			\textbf{b}, Examples of modes based on the axial mode (left) and the non-axial, V-shaped orbit (right). }
		\label{figureS1}
	\end{figure} 
	
\textbf{Passive cavity modes.} We calculate the cavity resonances with the COMSOL eigenfrequency solver module. The cavity resonances are the solutions of the scalar Helmholtz equation
\begin{equation} [\nabla^2 + k^2 n^2(x,y)] H_z(x,y) = 0 \end{equation}
with outgoing wave boundary conditions where $k$ is the free-space wave number and $H_z$ is the z-component of the magnetic field. The cavity length is $L = 20.0$ $\mu$m and the transverse width $W = L / \sqrt{2} = 14.1 $ $\mu$m, which is the maximum width for that the rectangle orbit is avoided (see Fig.~1c of the main text). For the fine-tuning of the cavity geometry, only the radius of curvature $R_c$ of the end mirrors is adjusted from $10~\mu$m (concentric, $g=-1$) to $20~\mu$m (confocal, $g=0$), while $L$ and $W$ are kept constant. The resonant modes are obtained in a spectral range centered at $\lambda_0=800$~nm, which is the approximate lasing emission wavelength in the experiments.
	
Figure~\ref{figureS1} shows the quality factors and wavelengths of high-$Q$ modes in concentric ($g=-1$), near-concentric ($g=-0.74$), and confocal ($g=0$) cavities. The numerically calculated mode wavelengths agree well with the analytic expression for the frequencies of Hermite-Gaussian modes~\cite{siegman1986lasers},
	\begin{equation}
    \nu_{m,q} = \frac{c}{2nL} \left[q + \frac{1}{\pi} \left(m+\frac{1}{2} \right)\arccos(g)\right]
    \label{eq:one},
    \end{equation}
where $\nu = c / \lambda$ is the frequency, $c$ is the speed of light, $n$ is the refractive index, $L$ is the cavity length, $g$ is the cavity stability parameter, and $(q, m)$ are the longitudinal and transverse mode numbers, respectively. The deviations between numerical and analytic mode wavelengths gradually grow as $m$ increases and reach $0.04\%$ for the highest-order high-$Q$ transverse mode ($m=23$) in the near-concentric cavity. The deviations are larger for the concentric cavity ($g=-1$), since it is at the border of the stability region where Eq.~(\ref{eq:one}) no longer holds. The fundamental transverse Hermite-Gaussian modes ($m=0$) have the highest $Q$-factors $Q_{max} \simeq 433$, which is equal to the $Q$-factor of a Fabry-Perot cavity with length $L$,
	\begin{equation}
    Q_{max} = \frac{2 \pi \nu n L}{c\ln(1/R)} 
    \label{eq:two}
    \end{equation}
where $\nu$ is the vacuum frequency and $R = [(1-n)/(1+n)]^2$ is the reflectivity of the cavity facet for normal incidence. The number of high-$Q$ modes shown in Fig.~2a is given by the number of modes whose $Q$-factor is above $0.8 \, Q_{max}$ (this range is marked by the horizontal dashed lines in Fig.~\ref{figureS1}).

In addition to the usual Hermite-Gaussian modes based on the axial orbit, modes based on V-shaped orbits (see Fig.~\ref{figureS1}) exist in confocal ($g=0$) and near-confocal ($g$ close to $0$) geometries. These modes are indicated as red squares in Fig.~\ref{figureS1}a, and in most cases exhibit higher $Q$-factors than the axial modes since the V-shaped orbit experiences total internal reflection at one mirror facet. The number of high-$Q$ modes in Fig.~2a refers only to the axial modes, excluding the non-axial orbits which are undesirable due to their non-directional output. \newline

  	% FIGURE 2
	\begin{figure}[b]
		\centering
		\includegraphics[width=\linewidth]{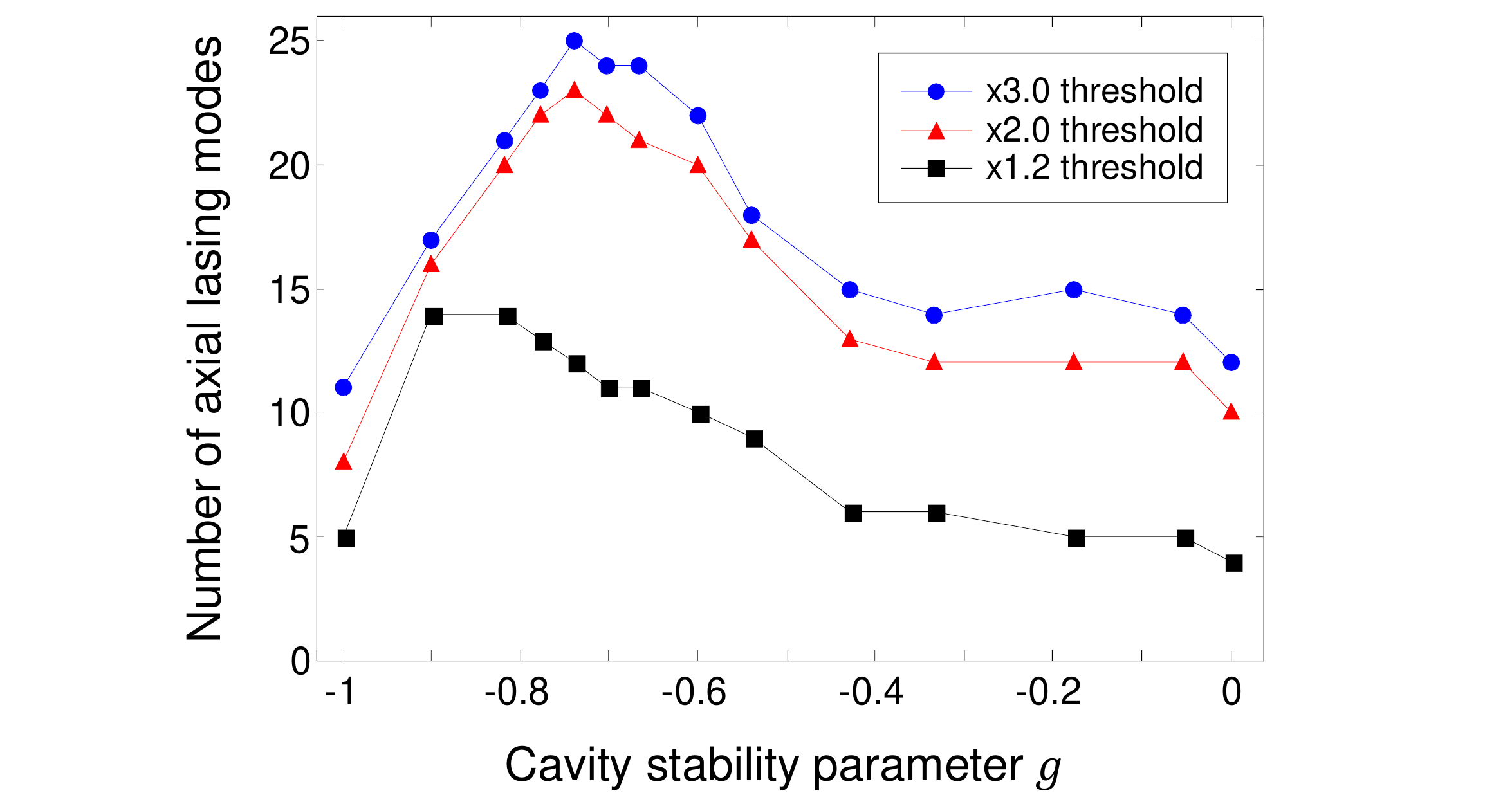}
		\caption{
			\textbf{Number of transverse lasing modes.} Numerically calculated number of different transverse lasing modes based on the axial orbit as a function of the cavity stability factor $g$ at three pumping levels.}
		\label{figureS2}
	\end{figure}   
	
\textbf{Gain competition.} We use the SALT (steady-state \textit{ab-initio} laser theory) to investigate the effect of mode competition in an active cavity~\cite{cerjan2016controlling}. We assume a spatially uniform distribution of pump and a flat gain spectrum. Both axial and non-axial modes are included in the simulations, and the presence of non-axial modes in the (near-) confocal cavities reduces the number of axial lasing modes, since the former have higher $Q$-factors than the latter. Figure~\ref{figureS2} shows the number of distinct transverse lasing modes at different pump levels, where only axial modes are counted. The pump level is defined with respect to the first lasing threshold of the axial modes. The maximum number of axial lasing modes is reached in the near-concentric regime of $g$ close to $-1$. The optimal $g$ value depends slightly on the pump level, but it remains at $g = -0.74$ when the pump exceeds twice of the lasing threshold. \newline

\textbf{Sample fabrication.} We use a commercial diode laser wafer (Q-Photonics QEWLD-808). The gain medium is a $12$~nm-thick GaAs quantum well, embedded in the middle of an undoped $400$~nm-thick Al$_{0.37}$Ga$_{0.63}$As guiding layer, which itself is between p-doped and n-doped Al$_{0.55}$Ga$_{0.45}$As cladding layers (each is $1.5~\mu$m thick). 
	
The laser cavities are fabricated by the following procedure. First the back contact made of Ni/Ge/Au/Ni/Au layers (thicknesses are $5/25/100/5/200$~nm, respectively) is deposited and thermally annealed at $390\,^{\circ}\mathrm{C}$ for $30$~s. Then a $300$~nm-thick SiO$_2$layer is deposited on the front side. The cavity shapes are defined by photolithography and transferred to the SiO$_2$ layer by reactive ion etching (RIE) with a CF$_{4}$ (30 sccm) and CHF$_{3}$ (30 sccm) mixture. After the removal of the photoresist, the remaining SiO$_2$ is used as mask for an inductively coupled plasma (ICP) dry etching with an Ar (5 sccm), Cl$_2$ (4 sccm), and BCl$_3$ (4.5 sccm) plasma mixture to create the cavities. The etch depth is about $4~\mu$m to etch all the way through the guiding layer and partially into the bottom cladding layer. After the ICP dry etching, the SiO$_2$ mask is removed by RIE and a buffered oxide etch (BOE).
	
The top metal contacts are defined by negative photolithography, followed by Ti/Au (thicknesses $20/200$~nm) deposition. The boundaries of the top contacts are $5~\mu$m away from the cavity edges to prevent the top contacts from hanging down and blocking the emission from the facets. The last process is the lift-off and the sample is cleaned by O$_2$ plasma afterwards. \newline

\textbf{Electrical pumping.} The fabricated samples are mounted on a copper plate and a tungsten needle (Quater Research, H-20242) is placed on the top contact for electric current injection. The lasers are pumped electrically by a diode driver (DEI Scientific, PCX-7401) which generates a series of rectangular current pulses. We use a pulse length of $2~\mu$s and a low repetition rate of $10$~Hz in order to reduce heating. \newline

  	% FIGURE 3
	\begin{figure}[t]
		\centering
		\includegraphics[width=\linewidth]{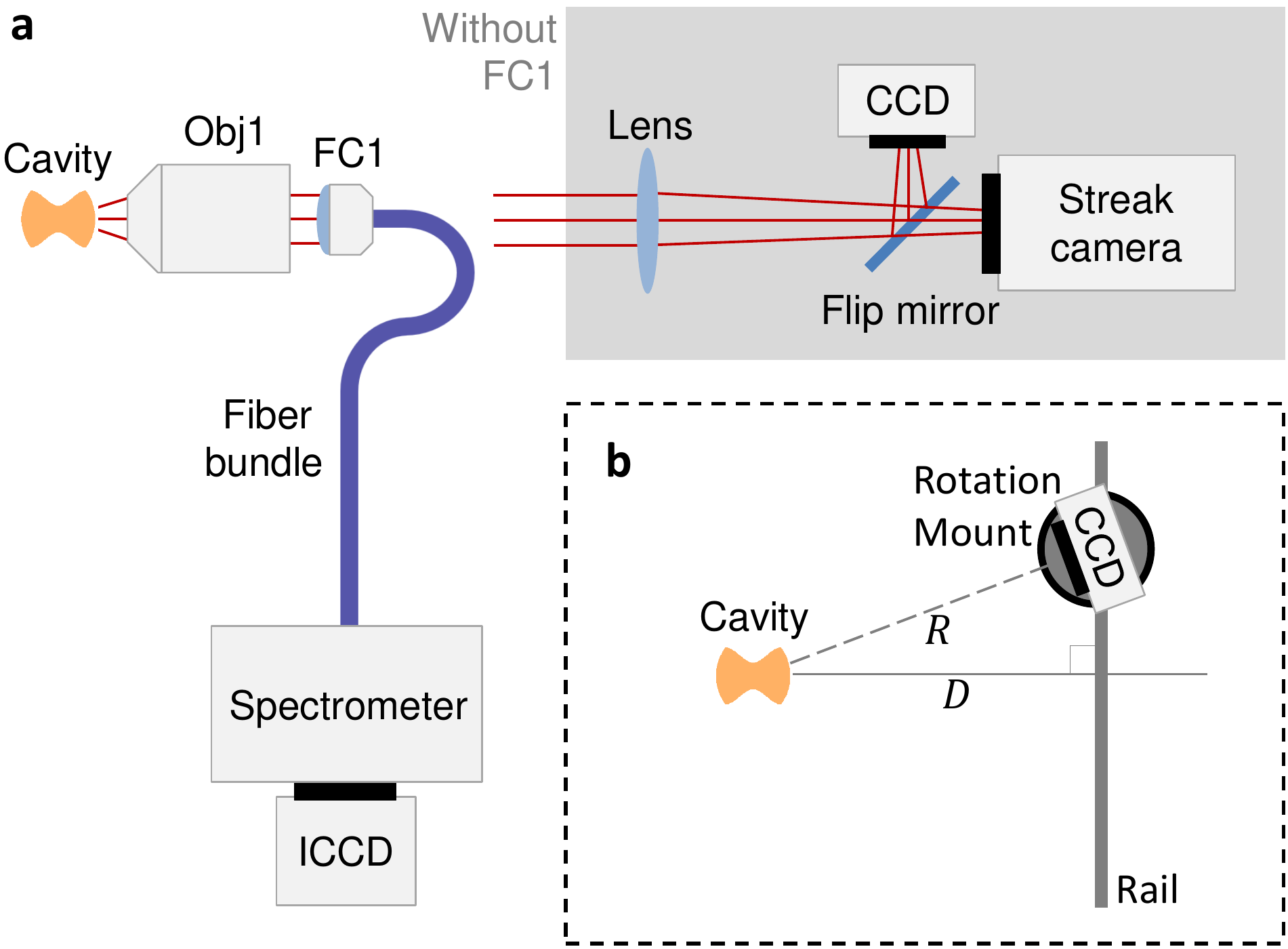}
		\caption{\textbf{Schematics of the experimental setup.}
			\textbf{a}, Optical setup to measure the emission spectrum. Emission from the end facet of a cavity is collected by an objective lens (Obj1) and coupled into a fiber bundle by a fiber collimator (FC1). For spatially resolved measurement of laser emission, the fiber collimator (FC1) is removed, and the end facet of the cavity is imaged onto a CCD camera or the entrance slit of a streak camera.
			\textbf{b}, Optical setup to measure the far-field emission patterns. The laser emission is directly measured with a CCD camera on a rotating mount on a rail.
			}
		\label{figureS3}
	\end{figure}   
		
\textbf{Laser testing.} The optical setups used for laser characterization are sketched in Fig.~\ref{figureS3}a. For spectral measurements, the laser emission is collected by a $20\times$ microscope objective (NA = 0.40). It is then coupled into a fiber bundle with a fiber collimator (NA = 0.50) behind the objective lens. Its spectrum is measured by an imaging monochromator (Acton SP300i) equipped with an intensified CCD camera (Andor iStar DH312T-18U-73). 
	
To test the laser directionality, we measure the far-field emission patterns with the setup in Fig.~\ref{figureS3}b. The objective lens is removed and the laser emission is measured after free-space propagation. A CCD camera (Allied Vision, Manta G-235B) is placed at a distance $D= 6$~cm away from the cavity. A large angular range is covered by moving the CCD camera on a rail while rotating the camera to face the cavity at every position. Since the distance $R$ from the cavity to the camera varies with the position, the measured intensity is rescaled by $1/R^2$ accordingly. The recorded images are stitched together in the horizontal direction and vertically integrated to obtain the far-field patterns shown in Fig.~3c. \newline
	
\textbf{Measurement of spatial coherence.} The spatial coherence of laser emission is measured using speckle patterns generated by a line diffuser. The microscopic structure of the line diffuser (RPC Photonics, EDL-20) consists of fine random elongated grains of about $10~\mu$m width on top of a quasi-periodic structure of $100~\mu$m scale.

The optical setup (Fig.~4a) consists of an objective lens that collects the laser emission and a line diffuser that is placed in the pupil plane of the objective ($6$~mm diameter). The laser emission fills the entire aperture of the objective and thus covers hundreds of random elongated grains of the line diffuser. A plano-convex lens in $f-f$ configuration between the diffuser and the CCD camera (Allied Vision, Mako G-125B) allows to measure the far-field speckle patterns. 
	
\begin{figure}[t]
		\centering
		\includegraphics[width=\linewidth]{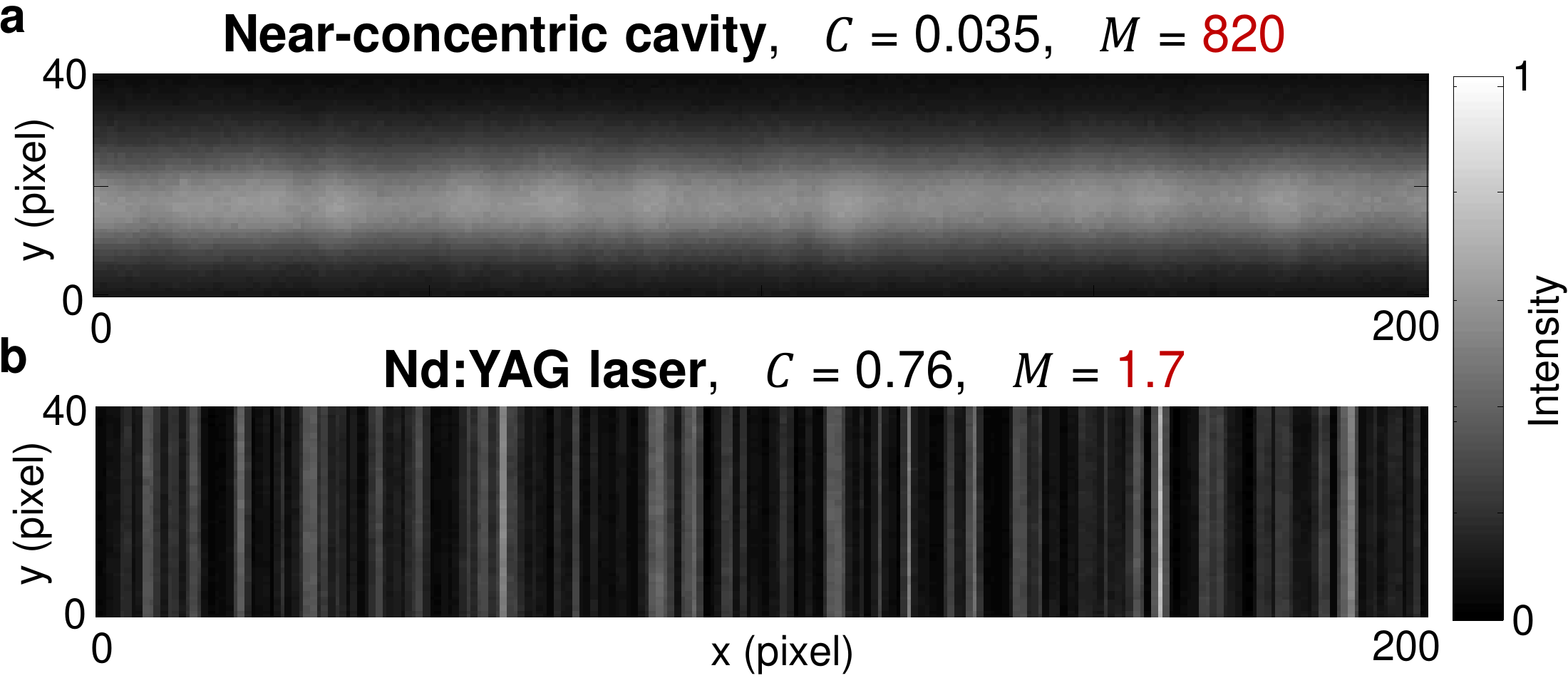}
		\caption{
			\textbf{Speckle patterns from a line diffuser.} 
			\textbf{a}, Laser emission from a near-concentric cavity ($g = -0.74, L=400~\mu$m) passes through a line differ, and the far-field speckle pattern has an intensity contrast $C = 0.035$. The effective number of distinct transverse lasing modes is $M = 1/C^2 = 820$.
			\textbf{b}, A spatially coherent Nd:YAG laser beam ($\lambda = 532$ nm) passing through the same diffuser creates a speckle pattern with contrast $C = 0.76$.}
		\label{figureS4}
	\end{figure}

Figure~\ref{figureS4}a is the measured speckle pattern from a near-concentric cavity laser. For comparison, the speckle pattern from a source with high spatial coherence, a frequency-doubled Nd:YAG laser (Continuum, Minilite), is also measured with the same optical configuration (Fig.~\ref{figureS4}b). The typical speckle size on the CCD camera is $2.5$~pixel calculated from the intensity autocorrelation function, so that the speckle contrast reduction due to undersampling is negligible~\cite{goodman2007speckle}. For each speckle contrast measurement, speckle patterns are repeatedly measured for different lateral positions of the line diffuser, and the speckle contrasts for these different disorder realizations are averaged.
    
In order to measure time-resolved speckle patterns, the CCD camera is replaced by a streak camera (Hamamatsu C5680) with a fast sweep unit (M5676). The streak camera is operated with $20$~ns-long time windows and the temporal resolution is about $60$~ps. The time-resolved speckle patterns in longer time windows are obtained by putting together multiple $20$~ns-long streak images for consecutive time intervals.
 
  	% FIGURE 4

 	% FIGURE 5
	\begin{figure}[t]
		\centering
		\includegraphics[width=\linewidth]{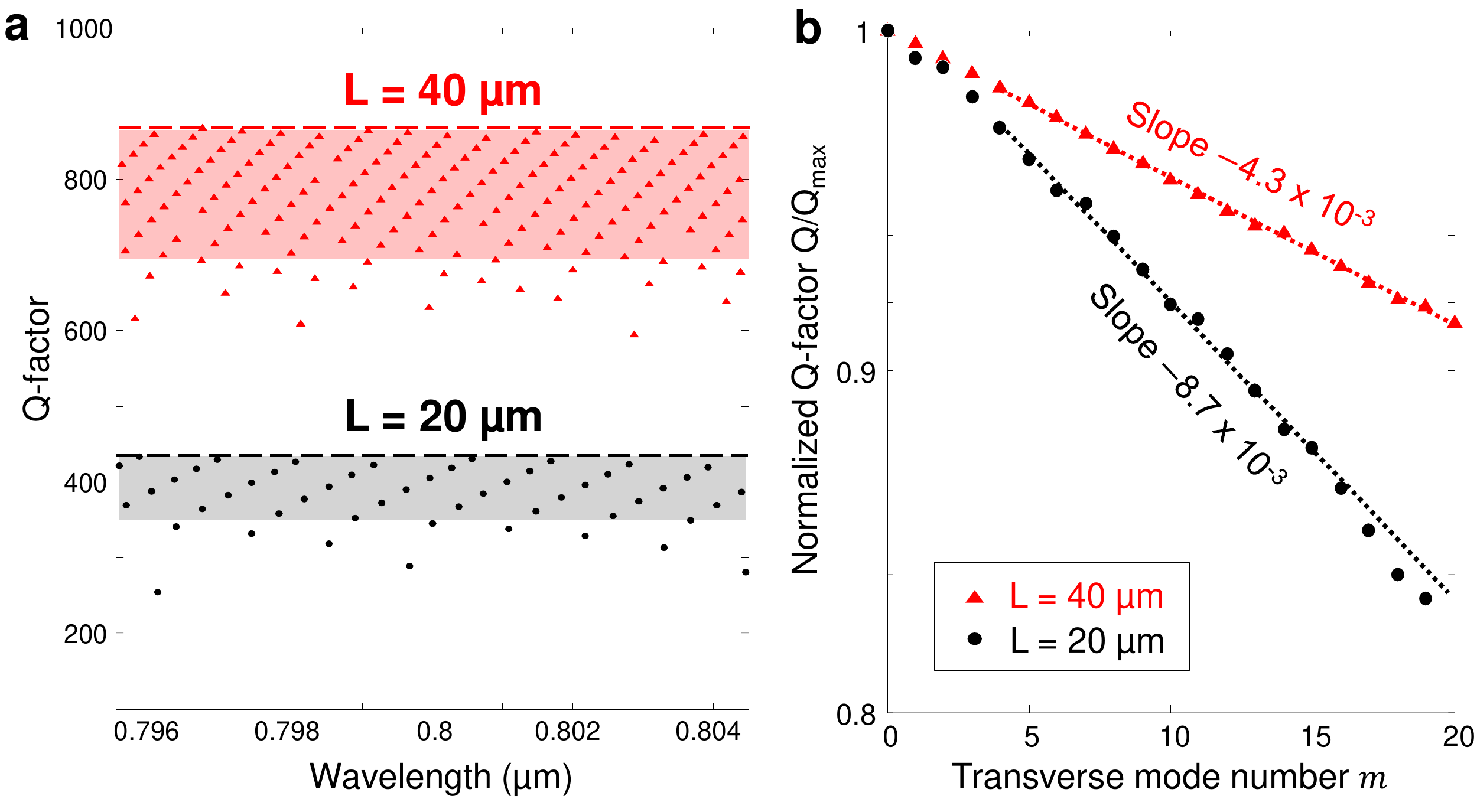}
		\caption{
			\textbf{Scaling of the number of transverse modes with cavity size.} 
			\textbf{a}, Calculated $Q$ factors and wavelengths of modes with relatively high $Q$ in near-concentric cavities ($g=-0.74$) with lengths $L=40~\mu$m (red triangle) and $L=20~\mu$m (black dots). The dashed lines represent the maximum $Q$-factors $Q_{max}$ in the two cavities. The shaded areas indicate the high-$Q$ regions of $0.8 \, Q_{max} < Q < Q_{max}$.
			\textbf{b}, Dependence of $Q$-factors on transverse mode number $m$ for $L=40~\mu$m (red triangle) and $L=20~\mu$m (black dots). $Q$ is normalized by $Q_{max}$ of each cavity. The dotted lines are linear fits and their slopes are shown.
			}
		\label{figureS5}
	\end{figure}

\subsection*{Scaling of the number of modes with the cavity size}
 	
When comparing different cavity geometries, we simulate small cavities with length $L = 20~\mu$m and width $W = 14~\mu$m to keep the computation time reasonably short. However, the cavities used in experiments are much larger with $L=400$--$800~\mu$m and $W=283$--$566~\mu$m, in order to increase the total number of transverse modes and lower the pump current density to reduce heating. To verify that the optimal value of the stability parameter $g$ found in simulations holds for larger cavities, we perform simulations with a $L=40~\mu$m-long cavity. 
    
Figure~\ref{figureS5} shows the high-$Q$ modes in two near-concentric cavities ($g=-0.74$) with $L=20~\mu$m and $L=40~\mu$m. The ratio $L/W = \sqrt{2}$ is the same for both cavities. The wavelengths and quality factors of the modes with relatively high $Q$ are shown in Fig.~\ref{figureS5}a. The maximum $Q$-factor $Q_{max}$ of the $L=40~\mu$m and $L=20~\mu$m cavities are $867$ and $433$, respectively, as expected from Eq.~(\ref{eq:two}). The shaded regions in Fig.~\ref{figureS5}a indicate the high-$Q$ regions of $0.8\,Q_{max} \leq Q \leq Q_{max}$. The high-$Q$ resonances of the $L=40~\mu$m-long cavity are more closely spaced than those of the $L=20~\mu$m-long cavity since the total number of resonances is proportional to the area of the cavity.
    
Figure~\ref{figureS5}b shows the dependence of the $Q$-factors on the transverse mode number $m$ for the near-concentric cavities ($g=-0.74$) for $L = 20~\mu$m and $L = 40~\mu$m. The $Q$-factors decrease approximately linearly with $m$, where the slope of the decrease for the $L = 40~\mu$m-long cavity is about one half of that for the $L=20~\mu$m-long cavity. This indicates that the number of different transverse modes with high-$Q$ is about twice as large for the $L=40~\mu$m-long cavity as for the $L=20~\mu$m-long cavity. This linear scaling is verified for another cavity geometry, $g=-0.54$, where the slopes are $1.2\times10^{-2}$ and $6.1\times10^{-3}$ for $L=20~\mu$m and $L=40~\mu$m-long cavities, respectively. Due to this linear scaling, the optimal value of $g$, at which the number of high-$Q$ transverse modes is maximal, is independent of the cavity size.

\subsection*{Frequency spacing of cavity resonances}
    
The first kink in Fig.~4e of the main text indicates that the speckle contrast saturates at the integration time of a few nanoseconds. This time scale is related to the average frequency spacing between neighboring modes. We estimate the average mode spacing using the simulation results for a small cavity and apply linear scaling with the cavity size as explained in the previous section. For the near-concentric cavity ($g=-0.74$) with $L=20$ $\mu$m, the number of high-$Q$ transverse modes is $23$ as given in Fig.~2a. The free-spectral-range (FSR) is given by the longitudinal mode spacing, $c/(2nL) = 2.225$~THz. Within one FSR, there is a series of transverse modes with $m = 0$--$22$. Thus the average frequency spacing between adjacent transverse modes is about $96.7$~GHz. 
    
In the experiments, the laser cavities with $L = 400~\mu$m have both $L$ and $W$ increased by a factor of $20$ compared to the simulated ones. Consequently, the FSR is reduced by a factor of 20, and the number of transverse modes within one FSR increases by 20. Therefore, the average mode spacing is reduced by a factor of $400$, which yields $242$~MHz. The beating of two modes is averaged out when the integration time is longer than the inverse mode spacing, which is about $4$~ns. This estimation gives the correct order of magnitude for the integration time at which the speckle contrast stops dropping in Fig. 4e.

\subsection*{Thermally-induced mode instability}	
 	
\begin{figure}[t]
		\centering
		\includegraphics[width=\linewidth]{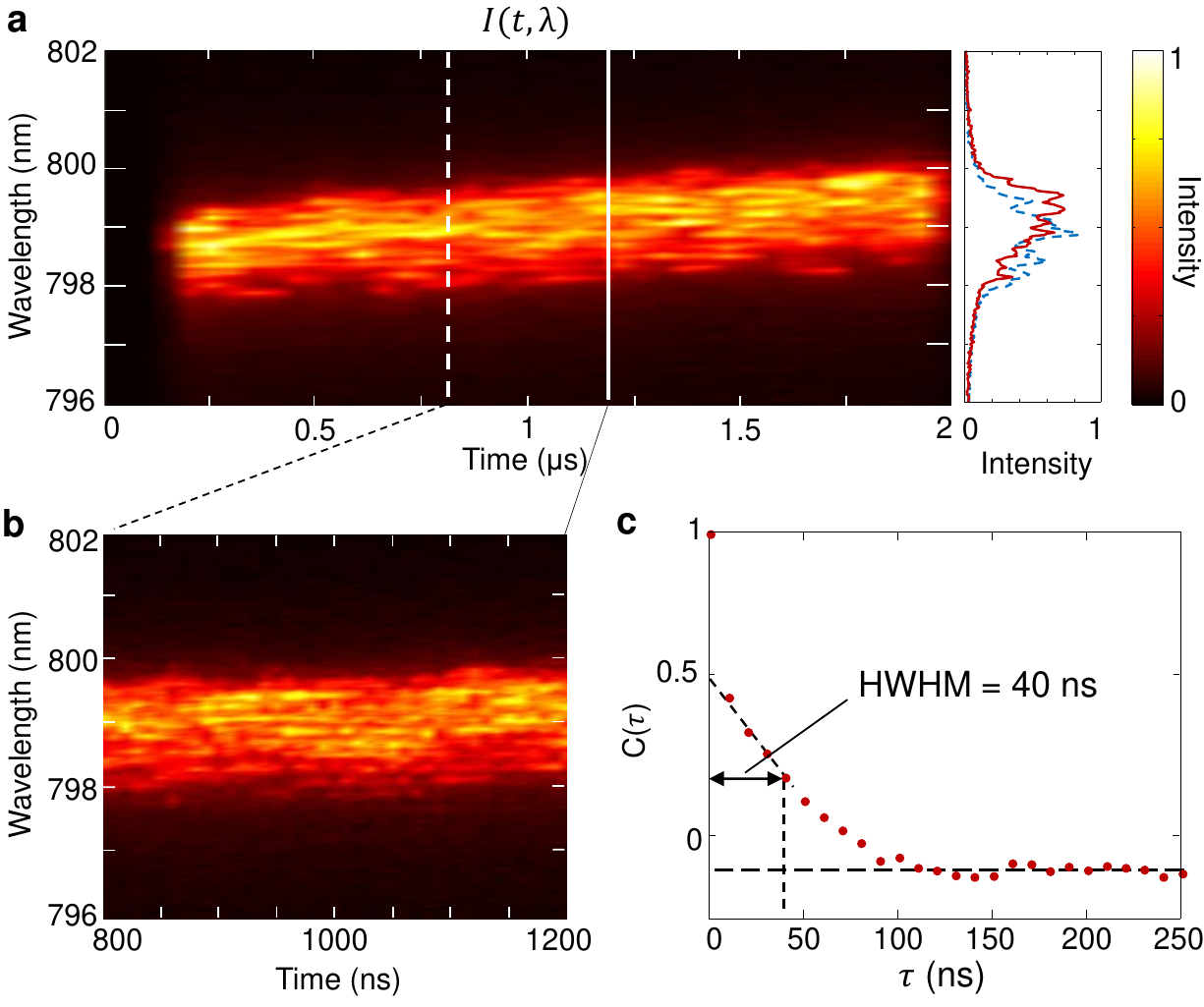}
		\caption{
			\textbf{Time-resolved spectra of laser emission.} 
			\textbf{a}, Measured spectrochronogram $I(t,\lambda)$ of laser emission from a near-concentric cavity ($g = -0.74, L=400~\mu$m) at the pump current of two times the lasing threshold. The temporal resolution is $50$~ns. The spectra at $0.8~\mu$s and $1.2~\mu$s are plotted on the right as dashed and solid lines, respectively. 
			\textbf{b}, Spectrochronogram during a $400$~ns long interval measured with $10$~ns resolution, showing that lasing peaks appear and disappear in time.
			\textbf{c}, The autocorrelation function $C(\tau)$ of the time-resolved spectra in \textbf{b}. The half width at half maximum (HWHM) of $C(\tau)$ is about $40$~ns.
			}
		\label{figureS6}
	\end{figure}

In Fig.~4e of the main text, the speckle contrast displays a second kink at several tens of nanoseconds after which the contrast further decreases. This behavior is caused by thermal effects that cause the lasing modes to change in time. We conduct time-resolved measurements of the lasing spectrum to observe the spectro-temporal dynamics. The gating function of the intensified CCD camera is used to acquire the lasing spectra with $10-50$~ns time resolution. The spectra from multiple measurements with consecutive time intervals during the pump pulse are combined to obtain the spectrochronogram of a whole pulse. Figure~\ref{figureS6}a is the measured spectrochronogram of a near-concentric cavity $(g=-0.74)$ laser with $L = 400~\mu$m. Since thermal equilibrium is not reached, the emission spectrum red-shifts during the pulse due to sample heating. Figure~\ref{figureS6}b is the spectrochronogram in a $400$~ns-long interval with a finer temporal resolution of $10$~ns. 
    
 	% FIGURE 6
	
The measured spectrochronogram reveals changes of the lasing spectrum during the pulse. Lasing peaks appear or disappear over the course of the pump pulse as different lasing modes turn on or off. In order to quantify the time scale of these changes, we calculate the temporal correlation function of the spectral changes defined as~\cite{bittner2018suppressing}
    \begin{equation}
    C(\tau) = \sum_{\lambda} \langle \delta I(t,\lambda) \delta I(t+\tau,\lambda) \rangle_t
    \label{eq:three},
    \end{equation}\\	
where $\delta I(t,\lambda) \equiv [I(t,\lambda)-\langle I(t,\lambda) \rangle_t] / \sigma_I(\lambda)$ is the normalized change of the emission intensity and $\sigma_I(\lambda)$ is the standard deviation. The half width at half maximum (HWHM) of the temporal correlation function $C(\tau)$ gives the time scale of the spectral dynamics. In Fig.~\ref{figureS6}c, the sharp drop of $C(\tau)$ at $\tau \sim 0$ is caused by the measurement noise. The HWHM of $C(\tau)$ is about $40$~ns, extrapolated from the more gradual decrease of $C(\tau)$ after the initial drop. This is approximately the integration time in Fig.~4e where the second kink occurs, thus the further reduction of the speckle contrast is caused by the switching of lasing modes. The new lasing modes generate distinct speckle patterns that are superposed to the ones created by the old lasing modes, reducing the intensity contrast of the time-integrated speckle patterns.

% Reference

%\bibliography{Reference} % Produces the bibliography via BibTeX.

\end{document}